\documentclass[%
reprint,
superscriptaddress,
citeautoscript,
showpacs,
 amsmath,amssymb,
 aps,
 prb,
floatfix,
longbibliography,
10pt
]{revtex4-1}

\usepackage{graphicx}

\usepackage{amsmath}

\newcommand\T{\rule{0pt}{2.6ex}}
\newcommand\B{\rule[-1.2ex]{0pt}{0pt}}
\newcommand{\ra}[1]{\renewcommand{\arraystretch}{#1}}

\newcommand{\TiC}{{Ti$_3$C$_2$T$_2$}}
\newcommand{\VC}{{V$_2$CT$_2$}}
\usepackage{bm}

\usepackage{color}

\begin{document}

\preprint{APS/123-QED}

\title{Effect of mixed surface terminations on the structural and electrochemical properties of two-dimensional \TiC\ and \VC\ MXenes multilayers}
\author{Nuala M. Caffrey}
\affiliation{School of Physics and CRANN, Trinity College, Dublin 2, Ireland}

\date{\today}

\begin{abstract}

MXenes, a family of layered transition metal carbides and nitrides, have shown great promise for use in emerging electrochemical energy storage devices, including batteries and supercapacitors. 
MXene surfaces are terminated by mixed --O, --F and --OH functional groups as a result of the chemical etching production process. 
These functional groups are known to be randomly distributed over the surfaces, with limited experimental control over their composition. 
There is considerable debate regarding the contribution of these functional groups to the properties of the underlying MXene material. For instance, their measured Li or Na capacity is far lower than that predicted by theoretical simulations, which generally assume uniformly terminated surfaces. The extent to which this structural simplification contributes to such discrepancies is unknown.
We address this issue by employing first-principles calculations to compare the structural, electronic and electrochemical properties of two common MXenes, namely \TiC\ and \VC, with both uniform terminating groups and explicitly mixed terminations. 
Weighted averages of uniformly-terminated layer properties are found to give excellent approximations to those of more realistic, randomly-terminated structures. This approximation holds for the lattice parameters, the electronic density of states and the work function. 
The sodium storage capacity and volume change during sodiation in the interlayer space of these MXenes with mixed surface terminations are also investigated. The redox reaction is shown to be confined to the terminating groups for low concentrations of intercalated Na, with the oxidation state of the metal atoms unaffected until higher concentrations of intercalated Na are achieved. 
Finally, the average open circuit voltage is shown to be very similar for both \TiC Na and \VC Na with mixed terminations, although it is highly sensitive to the particular composition of the terminating groups.

\end{abstract}

\maketitle


\section{Introduction}
Li-ion batteries (LIBs) currently dominate the high-performance electrochemical storage market, but limited Li resources mean that alternatives are being explored. Sodium-ion batteries (NIBs), in particular, could present a viable and cost effective replacement for LIBs due to the natural abundance and low-cost of sodium. Na has a similar redox potential to Li and exhibits similar intercalation chemistry. Unfortunately, current electrode materials for Li-ion batteries, such as graphite, show little electroactivity for NIBs~\cite{ge1988electrochemical}, necessitating the development of alternative electrodes. 

In this regard, the family of MXene materials show promise. MXenes are layered materials produced from their parent M$_{n+1}$AX$_n$ phases through a selective chemical etching of the covalently bonded A layer (M is an early transition metal, A is typically aluminum, X is either C or N and n = 1 -- 3).
Ti$_3$C$_2$ was the first MXene material to be experimentally produced, etched from its parent Ti$_3$AlC$_2$ phase using a hydrofluoric-acid treatment~\cite{naguib2011two}. 
Since then, at least 19 different MXene compositions have been synthesized with many more predicted to be stable \cite{khazaei2018inights}.

Hydrophilic, conductive, layered MXenes show promising electroactivity for Li-, Na-, and K-ions. 
\VC\ has a Li$^+$ capacity of 280~mAhg$^{-1}$ at 1C and 125~mAhg$^{-1}$ at 10C cycling rates~\cite{naguib2013new}. \TiC\ has a Na$^+$ capacity of 100~mAhg$^{-1}$ over 100 cycles~\cite{kajiyama2016sodium} and an initial K$^+$ capacity of 260~mAhg$^{-1}$ which is reduced to 45~mAhg$^{-1}$ after 120 cycles~\cite{xie2014prediction}. 
For comparison, conventional graphite cells for Li-ion batteries have a maximum specific capacity of 372~mAhg$^{-1}$, but are currently operating close to their theoretical limit and cannot handle high cycling rates~\cite{sivakkumar2010rate}. Given the number of MXene materials which remain to be investigated, there is considerable scope for improvement.

One means of improving the electrochemical behaviour of MXene materials is to tune their surface chemistry.
As a consequence of the chemical etching process, the surfaces of MXene layers are generally covered by mixed termination groups, predominately --O, --F and --OH, collectively designated T$_x$. These terminating groups are known to be randomly distributed over the surface with limited experimental control over their composition~\cite{naguib201425th, wang2015atomic, halim2016x}. 
Theoretical investigations into these materials generally make a significant simplification: they assume the MXene surface is covered by uniform terminations of --O, --F or --OH groups, rather than the randomly mixed terminations found experimentally.
They find that the specific terminations have a significant effect on the structural, electronic and electrochemical behaviour of MXene layers~\cite{dall2014high, tang2012mxenes, PhysRevB.87.235441}. As a result, there are large discrepancies between the calculated and experimental properties of these materials, particularly for quantities such as their capacity. 

In this work, first-principles density-functional theory (DFT) calculations are used to determine how the structural, electronic and electrochemical properties of two of the most popular members of the MXene family, namely \TiC\ and \VC, depend on the nature of the terminating groups. 
Given both the computational expense of modelling randomly populated surface terminations and the inability of the uniformly terminated structures to reproduce experimental results, we propose that a better approximation can be made by taking the weighted average of the properties of the uniformly terminated materials. This would allow much cheaper calculations as uniformly terminated structures require a smaller unit cell. Given the difficulties in achieving uniform terminations experimentally, we also discuss how some of the properties predicted for these structures are affected by random termination populations, particularly the ultra-low work function predicted for --OH terminated MXene layers. 
Furthermore, we investigate Na-ion intercalation and the charge storage mechanism of these structures with mixed terminations and show how sensitive the open circuit voltage is to the mixed terminations. 

This paper is organized as follows: Section~\ref{theory_methods} discusses the model for randomly mixed terminations while under the restriction of periodic boundary conditions as well as the choice of termination composition. Section~\ref{section:monolayers} and Section~\ref{section:stacked} show how the structural and electronic properties of both monolayer and multilayer \TiC\ and \VC\ MXenes differ from those which assume uniform terminations and compare the results to available experimental data. Finally, section~\ref{section:sodium} discusses the intercalation of Na into these layers.


\section{Theoretical Methods}\label{theory_methods}

\subsection{Modelling mixed terminations}

In order to model the effect of mixed terminations on a MXene layer, we randomly populate the surface sites of a $3 \times 3$ supercell of Ti$_3$C$_2$ and V$_2$C with --O, --F and --OH groups such that, when averaged over nine different supercells, they reproduce a typical ratio of surface terminations found experimentally. 
For \TiC, the terminating atoms are placed at the hollow sites directly above the central Ti atom, while for \VC, the terminating atoms are placed directly above the V atom on the opposite surface. This is in agreement with previous calculations~\cite{wang2015atomic, hu2014investigations}. Note that each layer has 18 possible binding sites -- 9 on each surface. A full structural relaxation is then performed.

\begin{figure}[ht!]
\begin{centering}
\includegraphics[width=1\linewidth]{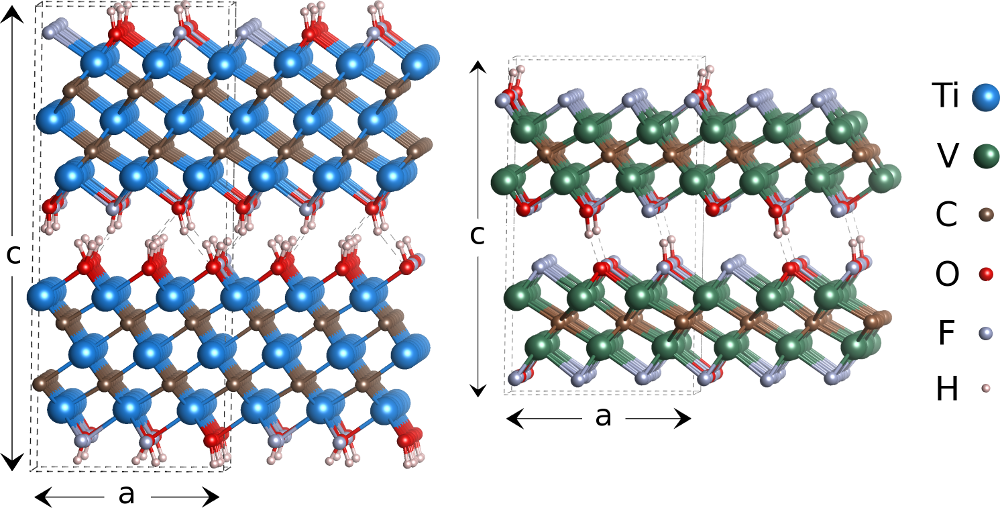}
\caption{\label{fig:clean_stacked_structures}(color online) Typical examples of stacked \TiC\ (left) and \VC\ (right) multilayers with randomly populated terminations and an AB-stacking configuration. The high percentage of --OH (--F) groups in such a typical mixed \TiC\ (\VC) structure is evident.}
\end{centering}
\end{figure}

The bulk stacked structure is modelled by randomly combining two of these monolayers in an AB-stacking configuration, as shown in Fig.~\ref{fig:clean_stacked_structures}. A further full structural relaxation is performed on these stacked layers to determine the interlayer spacing. Four different stacked structures are considered and their properties are averaged.

We define a simple weighted average of the properties of the three uniformly covered MXenes as:  

\begin{equation}
 \bar{P} = \frac{x}{2} P_{ \mathrm{-O} } + \frac{y}{2} P_{  \mathrm{-F} } + \frac{z}{2} P_{  \mathrm{-OH}  }
 \label{eqn:wa}
\end{equation}
The properties considered include, for example, the lattice constants, the density of states and the work functions. $P_{ \mathrm{-T} }$  refers to the value of the property associated with the uniformly terminated structures, and $x$, $y$ and $z$ are related to the populations of the --O, --F and --OH groups on the surface, respectively (c.f.~Table~\ref{tab:stoichiometry}).

\subsection{Termination composition}

Experiment has shown conclusively that the ratio of terminating groups on the surface of MXene layers depends critically on the details of the chemical etching process. Using a 10\% concentration of hydrofluoric acid (HF) to etch Ti$_3$AlC$_2$ resulted in ordered crystalline layers with AB-stacking and a stoichiometric formula of Ti$_3$C$_2$O$_{0.13}$F$_{0.83}$(OH)$_{1.04}$~\cite{wang2015resolving}. This composition is modelled in the results presented here. \footnote{Given the limited number of structures which we can work with, our model reproduces this as Ti$_3$C$_2$O$_{0.11}$F$_{0.864}$(OH)$_{1.025}$.} Etching the same MAX phase material with NH$_4$HF$_2$ produced layered Ti$_3$C$_{1.8}$O$_{2.3}$F$_{0.2}$~\cite{karlsson2015atomically}.
There is less experimental information available about the distribution and relative populations of terminating groups on \VC\ layers. To reflect the higher concentration of --F groups generally found in \VC\ compared to \TiC\ as reported by experiment~\cite{naguib2013new, harris2015direct}, we employ an average composition of V$_2$CO$_{0.42}$(OH)$_{0.4}$F$_{1.19}$ in the calculations presented here.
The chosen surface terminations of both materials are summarized in Table~\ref{tab:stoichiometry}.

\begin{table}[h!]\centering
\ra{1.2}
\setlength{\tabcolsep}{6pt} 
\begin{tabular}{lccc}
\hline \hline 
 & $\bm{x}$  & $\bm{y}$ & $\bm{z}$ \\
\cline{1-4}
Ti$_3$C$_2$O$_x$F$_y$(OH)$_z$ \T \B & 0.11 (\,6\,\%) & 0.86 (43\%) & 1.03 (51\%) \\
V$_2$CO$_x$F$_y$(OH)$_z$ \T \B & 0.42 (21\%) & 1.19 (59\%) & 0.40 (20\%) \\
\cline{1-4}
\hline
\end{tabular}
\caption{\label{tab:stoichiometry}\,The composition of terminating groups in \TiC\ and \VC\ as used in this work. They were chosen to reflect typical experimental values. The percentage contribution of each terminating group to the total is also given.}
\end{table}

\subsection{Density Functional Theory}

Density functional theory calculations are performed using the {\sc vasp} code \cite{Kresse1996, Kresse1999, PhysRevB.50.17953}. The Perdew-Burke-Ernzerhof (PBE) \cite{Perdew1996} parametrization of the generalized gradient approximation (GGA) is employed. The plane wave basis set is converged using an 800~eV energy cutoff. Structural relaxations of the monolayer supercell are carried out using a $3\times3\times1$ $k$-point Monkhorst-Pack mesh \cite{PhysRevB.13.5188} to sample the three-dimensional Brillouin zone, while the stacked layer supercell is relaxed using a $3\times3\times2$ $k$-point mesh. The electronic structure is then calculated using a $15\times15\times1$ and $8\times8\times4$ mesh for the monolayers and stacked layers, respectively.
Dispersion forces are included using the semi-empirical approach of Grimme (DFT-D3)~\cite{Grimme2006, Bucko2010}.
For calculations involving monolayers, a vacuum layer of at least 25~\AA\ is included in the direction normal to the surface to electronically decouple neighboring slabs and the dipole correction is applied~\cite{PhysRevB.46.16067, PhysRevB.51.4014}.
The work function, $\Phi = e V_{\mathrm{vacuum}} - E_{\mathrm{F}}$, is obtained by calculating the planar average of the electrostatic potential across the super-cell and taking the vacuum potential, $V_{\mathrm{vacuum}}$, sufficiently far from the surface along the surface normal direction.


\section{Results \& Discussion}

\subsection{Monolayers}\label{section:monolayers}

\subsubsection{Structural Properties}

The in-plane lattice constants of \TiC\ and \VC\ monolayers are given in Table~\ref{tab:monolayer_structure}. 
The results presented for the uniformly terminated monolayers are in good agreement with previous calculations in the literature~\cite{khazaei2013novel, bai2016dependence, eames2014ion}, showing that the --O terminated structures have the smallest in-plane lattice constant while the --OH terminated structures have the largest lattice constant. This is true for both \TiC\ and \VC.

\begin{table}\centering
\ra{1}
\setlength{\tabcolsep}{11pt} 
\begin{tabular}{llcc}
\hline \hline
 & & \textbf{\TiC} & \textbf{\VC}\\
\cline{1-4}
Uniform & & & \\
 & --O  & 9.13 & 8.63   \\
 & --F  & 9.24 &  8.90   \\
 & --OH & 9.26 &  8.93  \\
Mixed & & & \\
 & Random   & 9.24 &     8.84  \\
 & Weighted ($\bar{a}$)  & 9.24 &    8.89  \\
\hline
\end{tabular}
\caption{\label{tab:monolayer_structure}\,In-plane lattice parameters, in \AA, of monolayer \TiC\ and \VC, including those structures with uniformly terminated --O, --F and --OH groups, the average in-plane lattice constant of 9 different structures with randomly populated mixed terminations (according to the compositions given in Table~\ref{tab:stoichiometry}) and the lattice constants given by taking a weighted average of the lattice constants of the uniformly covered surfaces, calculated according to Eqn.~\ref{eqn:wa}. In all cases, the lattice constant is given for a $3 \times 3$ unit cell.}
\end{table}

The average in-plane lattice constant of the nine different \TiC\  monolayers with randomly populated terminations is 9.24~\AA, with individual structures having values ranging between 9.22~\AA\ and 9.27~\AA. Notably, this averaged value is identical to $\bar{a}$, the value achieved by taking a weighted average of the lattice constant of the three uniformly terminated monolayers (c.f. Eqn.~\ref{eqn:wa}). These values are in good agreement with an experimentally reported value of 9.17~\AA~\cite{mashtalir2013intercalation}.
The average in-plane lattice constant of \VC\ structures which explicitly considers mixed terminations is 8.84~\AA, with individual structures having values ranging between 8.79~\AA\ and 8.87~\AA. This averaged value is within 0.6\% of that found by taking the weighted average of lattice constants of the the uniformly terminated structures, 8.89~\AA.

It is clear that taking a simple weighted average of the uniformly terminated in-plane lattice constants gives excellent agreement with those determined by explicitly considering mixed terminations. We now consider if this approximation is also applicable to the electronic properties of these monolayers. 

\subsubsection{Electronic Properties}

\begin{figure}[ht!]
\begin{centering}
\includegraphics[width=0.95\linewidth]{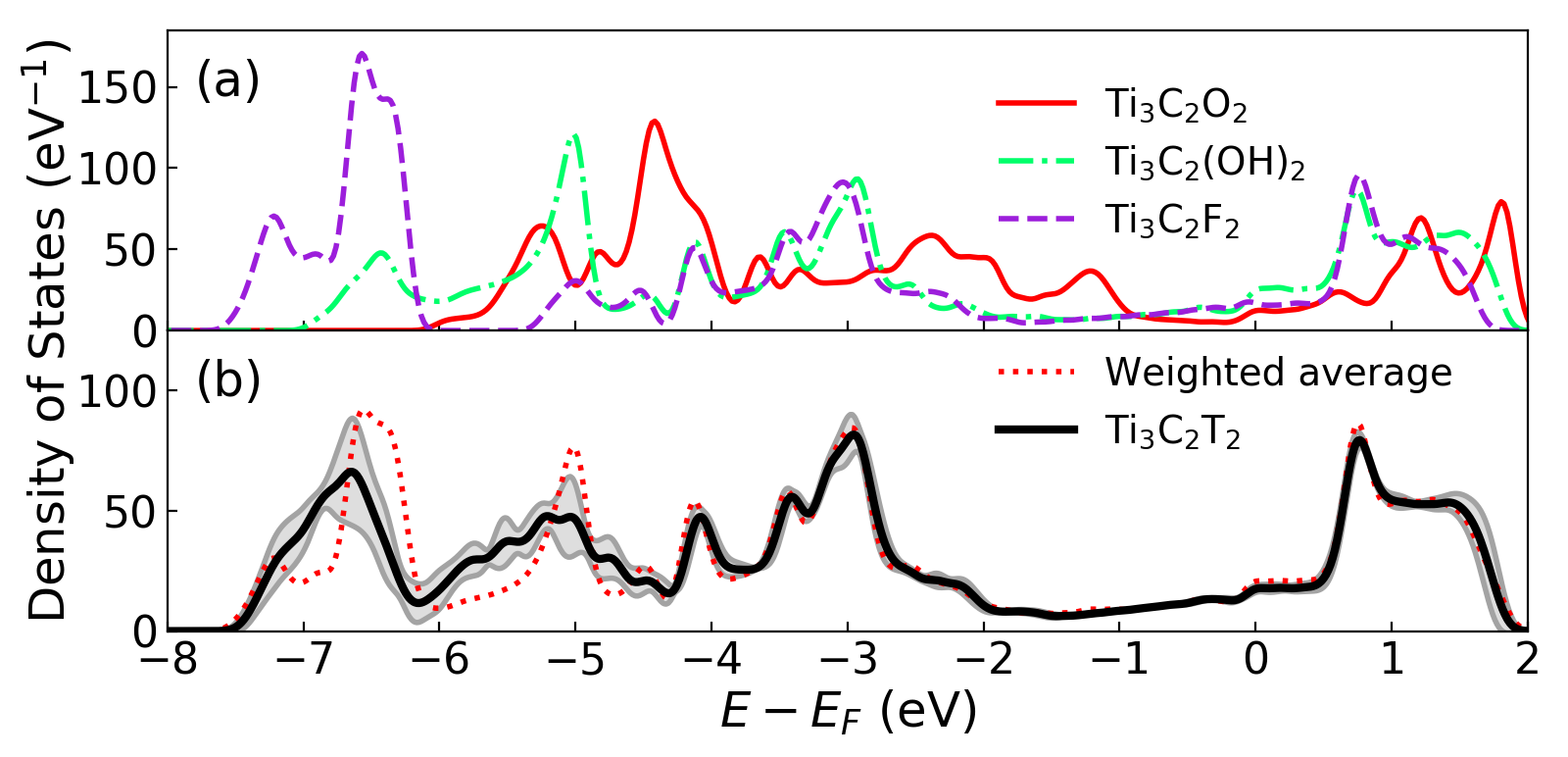}
\includegraphics[width=0.95\linewidth]{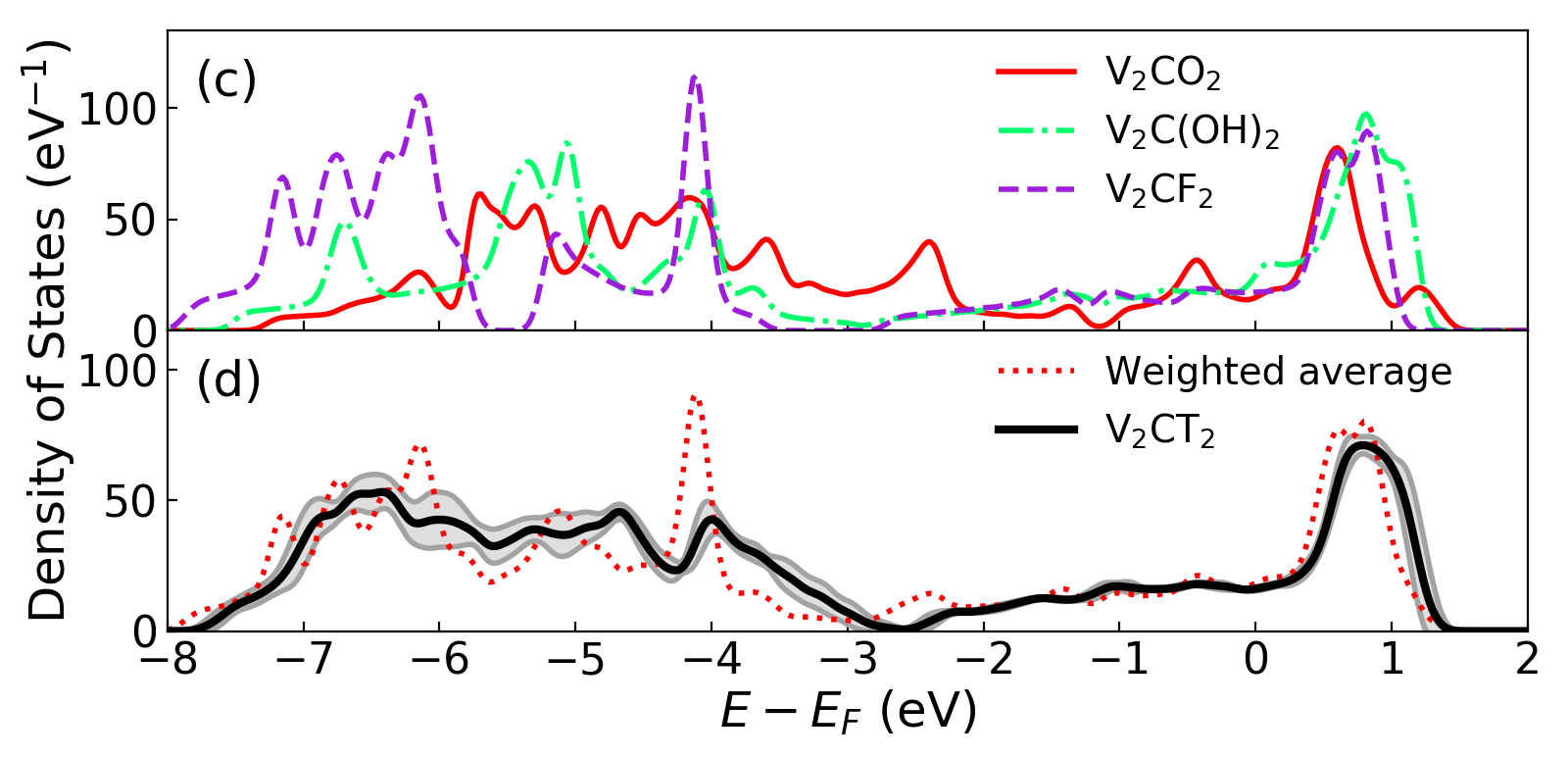}
\includegraphics[width=0.95\linewidth]{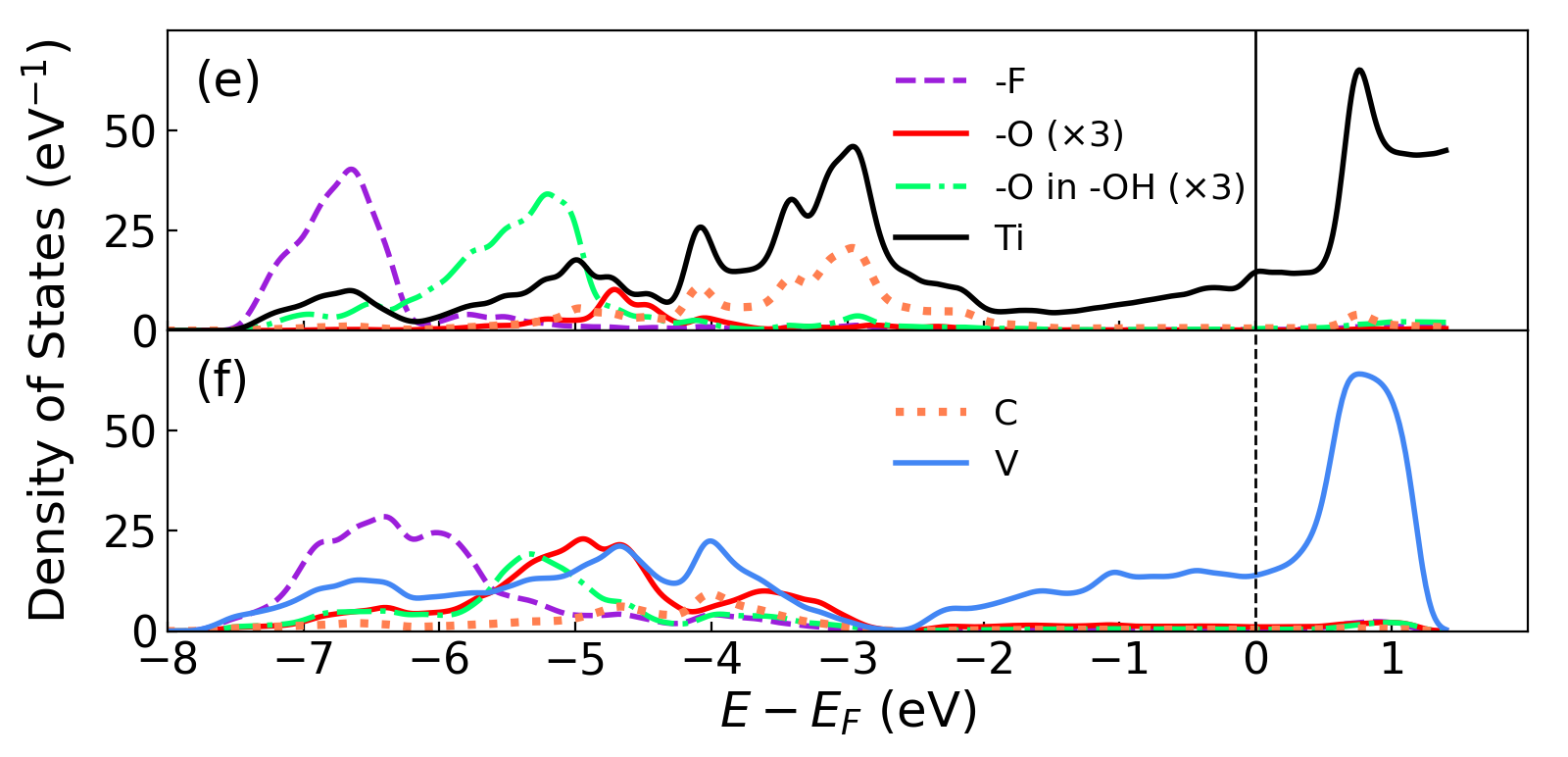}
\caption{\label{fig:averaged_dos} Total density of states calculated for (a) uniformly terminated Ti$_3$C$_2$ monolayers, (b) \TiC\ with randomly terminated \TiC\ monolayers (black solid line) and a weighted average of the uniformly terminated monolayers (red dashed line), (c) uniformly terminated V$_2$C monolayers, (d)  \VC\ with randomly terminated monolayers (black solid line) and a weighted average of the uniformly terminated V$_2$C monolayers (red dashed line). The gray shaded area in (b) and (d) show the population standard deviation of the individual structures with mixed terminations. The total projected density of states of (e) randomly terminated Ti$_3$CT$_2$ monolayers and (f) randomly terminated \VC\ monolayers.}
\end{centering}
\end{figure}

The total density of states (DOS) calculated for Ti$_3$C$_2$ and V$_2$C monolayers with uniform terminations are shown in Fig.~\ref{fig:averaged_dos}(a) and (c), respectively. They are in good agreement with previous calculations~\cite{yu2016prediction, champagne2017electronic}. 
Fig.~\ref{fig:averaged_dos}(b) and (d) then compare the averaged DOS of the structures with randomly populated terminations (solid black line) with those calculated as a weighted average of the uniformly terminated structures, shown with a red dotted line. 

First, we note that in all cases, and for both \TiC\ and \VC, the monolayers are found to be metallic, in agreement with experiment~\cite{mariano2016solution, champagne2017electronic}. 
Second, the DOS of the randomly populated monolayers is in excellent agreement with that calculated as a weighted average in a wide energy range between $-4$~eV and $+2$~eV ($-2$~eV and $+2$~eV) for \TiC\ (\VC). 
The origin of this agreement lies in the identity of the states located around the Fermi level of both materials. This can be seen by considering the density of states projected onto the relevant atomic orbitals, shown in Fig.~\ref{fig:averaged_dos}(e) and (f). The density of states of \TiC\ between $-4$~eV and $+2$~eV is predominately comprised of contributions from Ti and C. For \VC, contributions from V and C dominate in the energy region between $-2$~eV and $+2$~eV. States associated with the terminating groups are located lower in energy. 

As a result, in the energy window where states from the MXene layer, rather than the terminating atoms, dominate, the density of states of the structures with randomly populated terminations can be reproduced by a weighted average, as shown in Fig.~\ref{fig:averaged_dos}(b) and (d).  
For energies outside this region, the detailed agreement is not as good, suggesting that the exact contribution of the terminating groups cannot be determined by a weighted average of the individual components. The origin of this behaviour is related to the mutual interactions between the terminating groups. 
For example, the large peak in the DOS of V$_2$CF$_2$ at $-4$.2~eV (c.f. Fig.~\ref{fig:averaged_dos}(c)), which due to a hybridization between F, C and V states, is not present for the case of randomly mixed terminations, despite their high concentration (59\% of the total terminations).

\subsubsection{Work Function}

It has been predicted that MXene monolayers which are uniformly terminated with --OH groups will have an ultra-low work function of 1.85~eV, allowing them to function as very efficient field emitters~\cite{khazaei2015oh}.
However, given the difficulties associated with experimentally controlling the surface terminations of these materials, and the intimate relationship between a material's work function and the particulars of its surface, it is not clear that this ultra-low work function can be realistically achieved. 
Here, we determine whether the low work-function associated with uniform --OH terminations persists even when other terminating groups are present. 

\begin{figure}[htp]
\begin{centering}
\includegraphics[width=\linewidth]{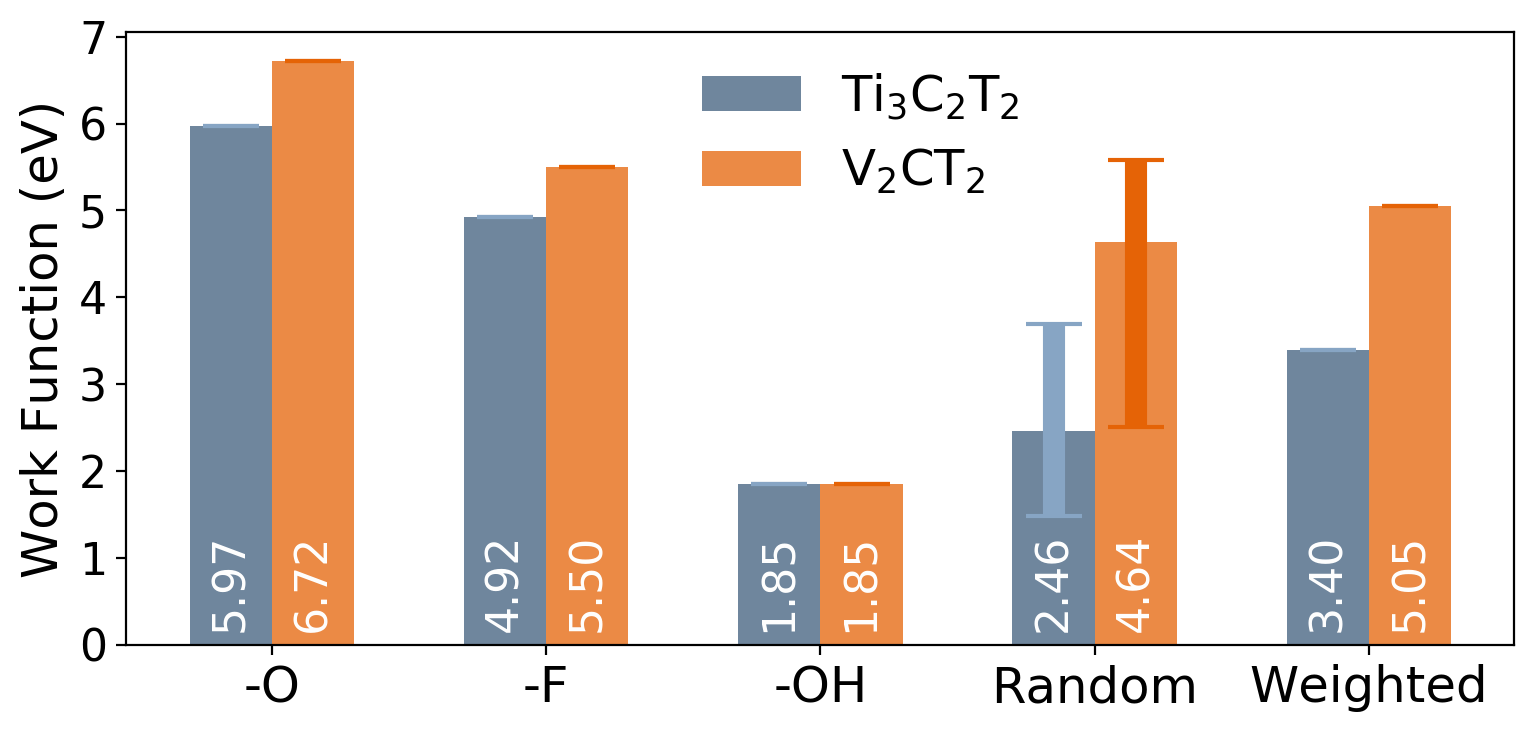}
\caption{\label{fig:work_function} Work functions of both uniformly and mixed terminated \TiC\ (blue bars) and \VC\ (orange bars). The error bars on the work-function calculated for the mixed terminations represent the range of values calculated for each individual structure. Note that the work functions of bare Ti$_3$C$_2$ and V$_2$C monolayers are 3.91~eV and 4.60~eV, respectively. }
\end{centering}
\end{figure} 

The work functions of the uniformly terminated structures, the average value of the randomly terminated structures as well as the value found by taking a weighted average of the uniformly terminated structures are shown in Fig.~\ref{fig:work_function}. The values for the uniformly terminated surfaces are in good agreement with previous calculations~\cite{khazaei2015oh, liu2016schottky}. In particular, we find a very high work function for structures with a uniform --O termination and an ultra-low work function for structures terminated with uniform --OH groups. The work functions of --O and --F terminated \VC\ are higher than those of \TiC, by 0.75~eV and 0.58~eV, respectively. 

For \TiC\ monolayers with mixed terminations, the averaged work function is 2.46~eV. Note that this was calculated as the averaged work function of 18 different surfaces; Individual surfaces had work functions ranging between 1.44~eV and 3.68~eV. For the \VC\ monolayers with mixed terminations, the average work function is 4.64~eV, with the work functions of individual surfaces ranging between 2.5~eV and 5.6~eV.

The general trend is clear: \TiC, which has a high percentage of terminating -OH groups (51\%), has a lower work-function than \VC\, which is dominated by -F groups (59\%). However, taking a simple weighted average of work functions of uniformly terminated structures results in a large error, particularly for \TiC.

\begin{figure}[htp]
\begin{centering}
\includegraphics[width=\linewidth]{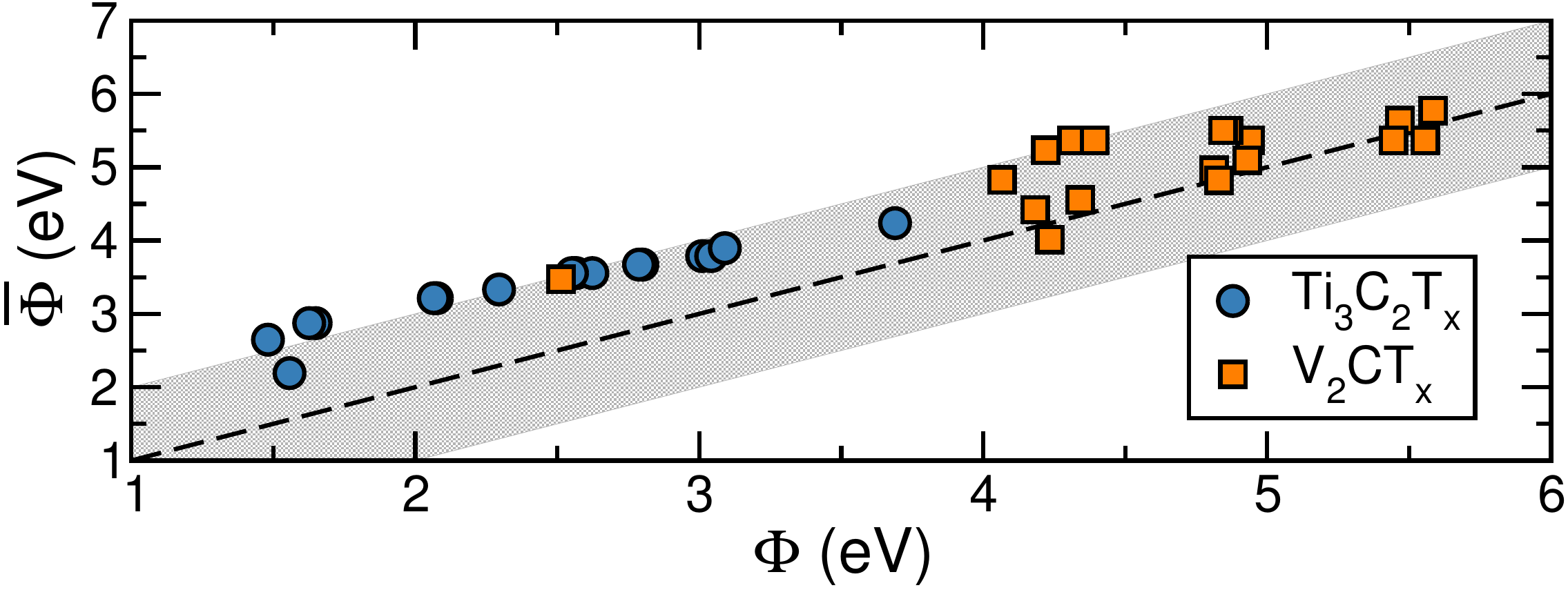}
\caption{\label{fig:wf_vs_wa} The work function calculated for each individual surface, $\Phi$, for \TiC\ (blue circles) and \VC\ (orange squares), compared to the value that would be found by taking the weighted average of the uniformly covered surfaces, $\bar{\Phi}$. The black dashed line corresponds to the case when $\Phi = \bar{\Phi}$, and the gray shaded area shows a region within $\pm 1$~eV of this line.}
\end{centering}
\end{figure} 

To probe this further, Fig.~\ref{fig:wf_vs_wa} compares the calculated work function of each of the eighteen individual surfaces, $\Phi$, to the value that would be found by taking the weighted average of the uniformly covered surfaces, $\bar{\Phi}$. Note that the composition of each individual surface was taken into account here -- it is the averaged composition of these eighteen surfaces that gives the composition reported in Table~\ref{tab:stoichiometry}. It is clear that $\bar{\Phi}$ always overestimates the actual work function of the \TiC\ surface by approximately 1~eV. On average, this results in an error of 38\%. For \VC, the average error is much smaller, at 9\%, although the error of some individual structures is larger. Agreement tends to improve when structures with high numbers of --OH groups are excluded, even if those groups are located on the opposite surface. Indeed, this is reason for the better agreement for \VC\ which, on average, has only 20\% --OH groups on the surface, compared to 51\% for \TiC. Finally, we note that even the maximum value of the \TiC\ work function found in these calculations (3.68~eV at 22\% --OH concentration) is significantly lower that the experimentally reported values of 5.28~eV~\cite{mariano2016solution} or 4.35~eV~\cite{kang2017mxene}. The composition of the terminating groups in those structures was not reported, however. It is likely that significant quantities of --O groups or H$_2$O molecules were present on their surfaces. 

\begin{figure}[htp]
\begin{centering}
\includegraphics[width=\linewidth]{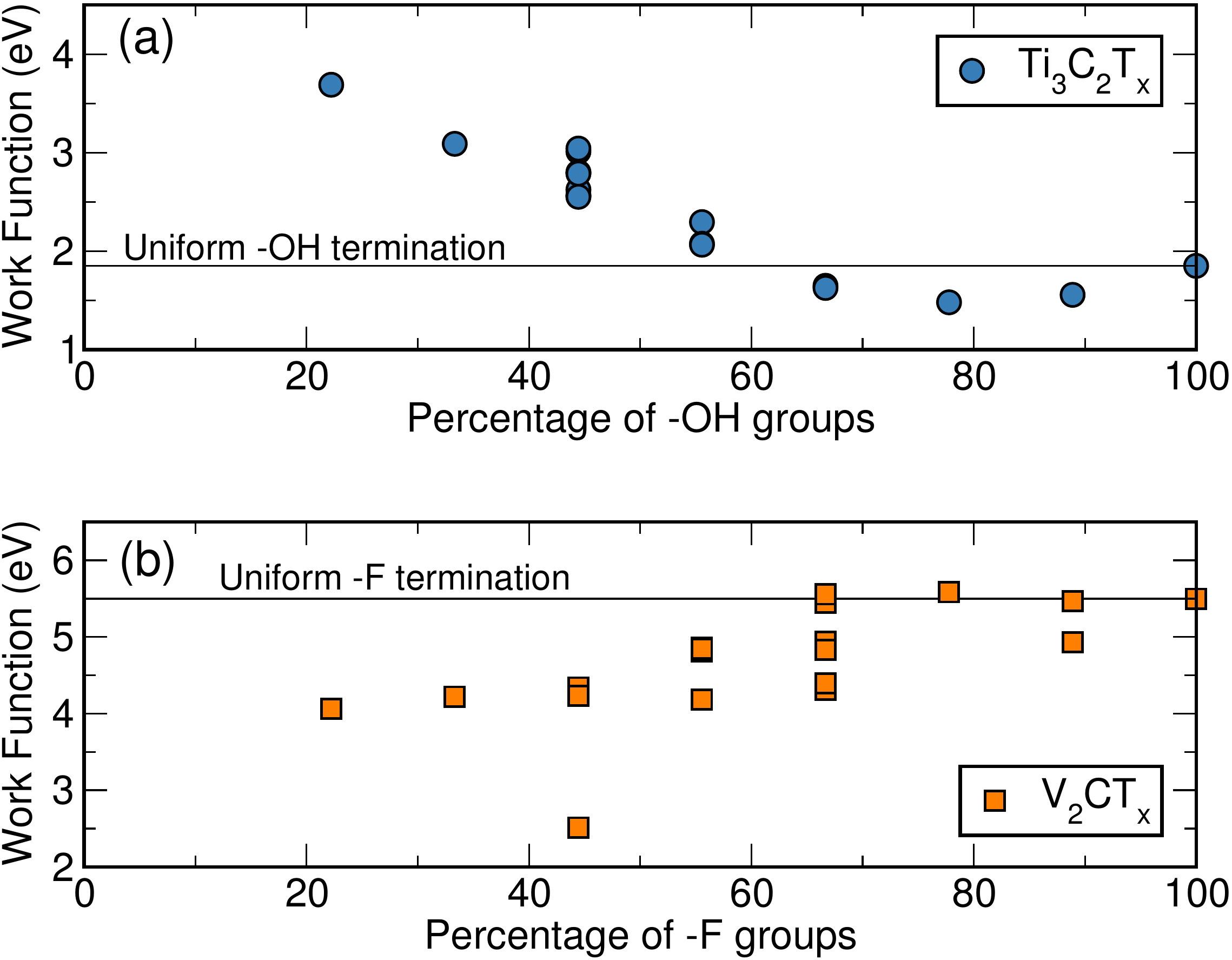}
\caption{\label{fig:oh_vs_wf} (a) The work function calculated for each individual \TiC\ surface, $\Phi$, as a function of the percentage of --OH groups in that particular structure. (b) The work function calculated for each individual \VC\ surface, $\Phi$, as a function of the percentage of --F groups in that particular structure. Note that for each concentration of --OH  or --F groups, the other two termination types are varying randomly - it is for this reason there is a spread of almost 0.5~eV for an --OH concentration of 44.4\%. }
\end{centering}
\end{figure}

Fig.~\ref{fig:oh_vs_wf}(a) then shows how the work function of the individual surfaces of \TiC\ with mixed terminations depends on the percentage of --OH groups on the surface. We find that approximately 60\% --OH groups is sufficient to achieve a work function of between 1.50~eV and 1.85~eV. 
Similarly, for \VC, there is a general trend for the work function to increase as the concentration of --F groups on the surface increases, as shown in Fig.~\ref{fig:oh_vs_wf}(b).

We can conclude that achieving very fine control over the surface terminations is not necessary in order to tune the work function to its extremal values. The ultra-low work function predicted for a uniform termination of --OH groups can also be achieved with approximately 60\% --OH groups. Similarly, work functions above 5~eV can be achieved with approximately 50\% --F groups on the surface. 

\subsection{Stacked Layers}\label{section:stacked}

The chemical etching process used to produced MXene materials usually results in stacked layers rather than completely delaminated layers~\cite{naguib2015large}. The interlayer space can then be intercalated by ions such as Li or Na or even small molecules~\cite{mashtalir2013intercalation, mashtalir2015amine}
Here, we discuss the structural and electronic properties of stacked \TiC\ and \VC\ with randomly populated terminations according to Table~\ref{tab:stoichiometry}.

\subsubsection{Structural Properties}

The calculated lattice constants of stacked \TiC\ and \VC\ are given in Table~\ref{tab:clean_structure}. 
\begin{table}\centering
\ra{1.4}
\setlength{\tabcolsep}{18pt} 
\begin{tabular}{llcc}
\hline \hline
 & & \textbf{a (\AA)} & \textbf{c (\AA)}\\
\cline{1-4}
\multicolumn{2}{l}{\TiC\ Uniform} &   & \\
 & --~O  & 9.06 & 18.72   \\
 & --~F  & 9.13 & 18.46  \\
 & --~OH & 9.27 & 19.86  \\
\multicolumn{2}{l}{\TiC\ Mixed} & & \\
 & Random   & 9.16 & 19.47   \\
 & Weighted & 9.20 & 19.20    \\
\multicolumn{2}{l}{\TiC\ Experiment} & & \\
 & Ref.~[\onlinecite{wang2015resolving}]   & 9.147 & 19.15   \\
\hline
\multicolumn{2}{l}{\VC\ Uniform}&   & \\
 & --~O  & 8.63 & 13.70    \\
 & --~F  & 8.90 & 14.07   \\
 & --~OH & 8.96 & 14.12   \\
\multicolumn{2}{l}{\VC\ Mixed}&   & \\
 & Random   & 8.84 & 14.07  \\
 & Weighted & 8.90 & 14.07  \\
\cline{1-4}
\cline{1-4}
\hline
\end{tabular}
\caption{\label{tab:clean_structure}\,Structural parameters of stacked \TiC\ and \VC, with uniform (--O, --OH and --F) and mixed terminations according to the composition given in Table~\ref{tab:stoichiometry}, for a $3 \times 3$ unit cell. The lattice constants found by taking a weighted average of the uniformly terminated values are also given. Note that Ref.~\onlinecite{wang2015resolving} has the same composition as considered in this work.}
\end{table}
Compared to the monolayers, we find that stacking has only a marginal effect on the in-plane lattice constants, reducing them by no more than 0.9\%. This is the case for both uniform and mixed terminations. The agreement between the in-plane lattice constant of the mixed termination stacked \TiC\ layers and the experimental value for a structure with the same composition is excellent, with values of 9.16~\AA\ and 9.147~\AA, respectively~\cite{wang2015resolving}.

The out-of-plane lattice constant, $c$, depends sensitively on the nature of the surface terminations. The $c$ lattice constant of Ti$_3$C$_2$O$_2$ is smallest at 18.72~\AA, and largest for --OH terminated structures at 19.86~\AA. These differ from the experimental value of 19.15~\AA\ by 2.2\% and $-3.7$\%, respectively~\cite{wang2015resolving}. Considering mixed terminations improves the agreement with experiment, giving a value of 19.47~\AA, an overestimation of only 1.7\%. Other experimental compositions report values of $c$ ranging between 19.3~\AA\ and 20.89~\AA~\cite{kajiyama2016sodium, osti2017influence, mashtalir2013intercalation, lukatskaya2013cation, shi2014structure, chang2013synthesis}. 

Here again, we find that taking the simple weighted average of the lattice constants of the uniformly terminated \TiC\ stacked layers results in excellent agreement with the values found by explicitly considering mixed terminations. The weighted in-plane lattice constant differs from the one calculated with randomly populated terminations by 0.4\%, while the out-of-plane lattice constant differs by 1.4\%. 

For the case of the \VC, the $c$ lattice constant is also strongly dependent on the composition of the surface terminations. It ranges from 13.70~\AA\ for --O terminations to 14.12~\AA\ for uniform --OH terminations. Due to the prevalence of --F terminations on the surface of the structures with randomly populated terminations, the $c$ value is found to be 14.07~\AA, identical to that of a structure with uniform --F terminations. 
As for \TiC, the in-plane lattice constants determined as a weighted average are in excellent agreement with those determined for the randomly terminated structures with differences of 0.7\% and 0.0\% for the in-plane and out-of-plane lattice constants, respectively. 

However, these values - whether uniform or mixed - are considerably smaller than any reported experimental $c$ values for \VC, which range between 19 and 24~\AA~\cite{naguib2013new, bak2017ion}. This is a considerable difference of 5 -- 10~\AA\ and as such cannot be attributed to a difference in the exact composition of the terminating groups. It is more likely that the experimental hypothesis that there are significant quantities of H$_2$O molecules intercalated between the layers of \VC\ is correct~\cite{naguib2013new, harris2015direct, ghidiu2016ion}. 

\subsection{Na Intercalation}\label{section:sodium}

To gain further insight into the effect of mixed terminations on the intercalation chemistry of these MXene layers, we now consider the structural and electronic changes that occur after Na intercalation in structures with randomly populated terminating groups. In the following, these intercalated structures are designated Ti$_3$C$_2$T$_2$Na$_x$ and V$_2$CT$_2$Na$_x$. A full intercalated Na monolayer corresponds to $x = 1$, or 9 intercalated Na atoms in the $3 \times 3$ unit cell considered here.

Compared to the unintercalated MXene layers, there are two changes to note. The first is related to the stacking. 
The un-intercalated structure has a Bernal-like AB-stacking, as shown in Fig.~\ref{fig:clean_stacked_structures}. After Na intercalation, the layers shift to a simple hexagonal stacking, in agreement with experiment~\cite{wang2015atomic}. In this configuration, the terminating atoms of adjacent layers are located directly opposite one another.
Furthermore, ion intercalation typically causes the number of --OH groups on the surface to be reduced or even eliminated~\cite{xie2014role, peng2014unique, xie2014prediction, yu2016prediction}. This instability of the H atoms in --OH terminations is also found to be the case here. As a result, we assume that after Na intercalation, each --OH group is replaced with a --O termination. 
Finally, in agreement with experiment~\cite{wang2015atomic} and previous calculations~\cite{tang2012mxenes, yu2016prediction}, the Na atoms are located directly on top of the carbon atoms for both \TiC\ and \VC.

\subsubsection{Lattice Constants}

\begin{figure}[ht!]
\begin{centering}
\includegraphics[width=\linewidth]{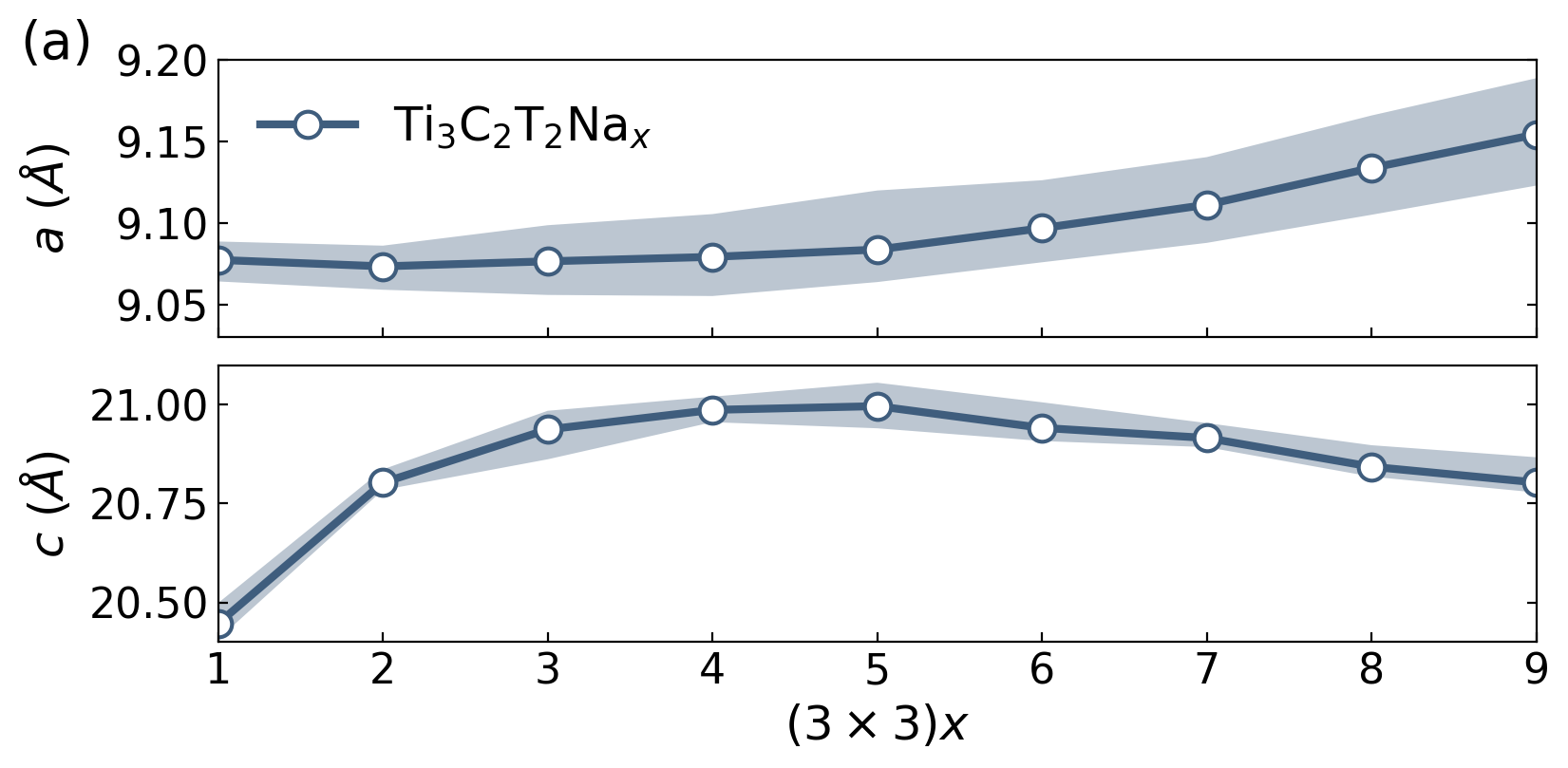}
\includegraphics[width=\linewidth]{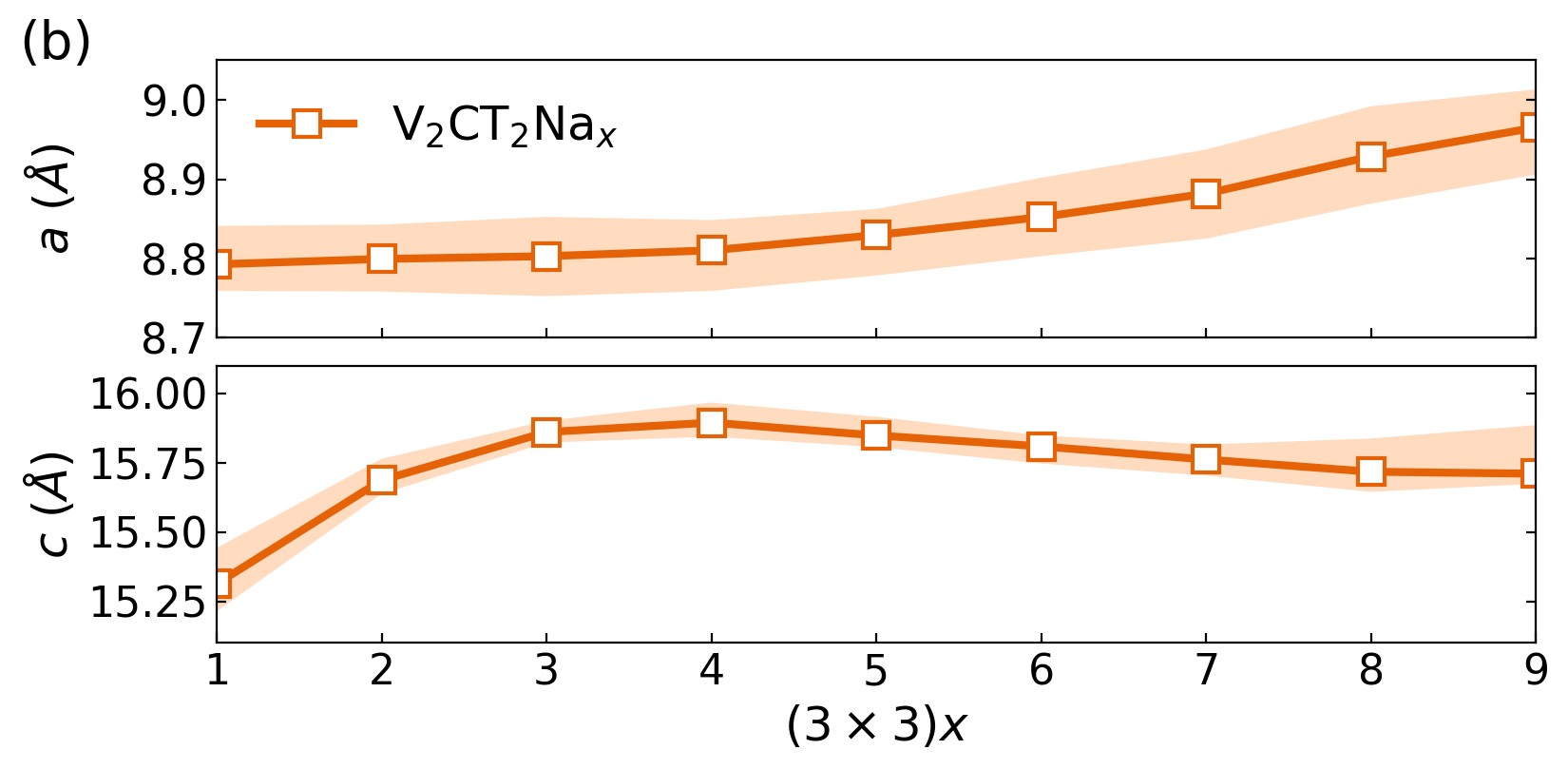}
\caption{\label{fig:structure_changes} Changes in the lattice parameters of a $3 \times 3$ supercell of stacked (a) Ti$_3$C$_2$T$_2$Na$_x$ and (b) V$_2$CT$_2$Na$_x$ as a function of the concentration of Na atoms, $x$. 
The solid lines are averages over 4 configurations with mixed terminations which have individual values that range between the shaded areas. }
\end{centering}
\end{figure}

The structural properties of \TiC\ and \VC\ with mixed terminations are shown in Fig.~\ref{fig:structure_changes} as a function of the number of intercalated Na atoms. 
For \TiC, we find that the in-plane lattice constant averages 9.09~\AA\ before intercalation for the structure with a ground state Bernal-like stacking and with all hydrogen atoms removed. After Na intercalation, the in-plane lattice constant initially decreases slightly by 0.02~\AA\ until approximately $x=2/9$. This agrees Shi et al.~\cite{shi2014structure}, who showed experimentally that the in-plane lattice constant decreases after Na intercalation by approximately 0.03~\AA\ (for a $3 \times 3$ unit cell). For concentrations greater than $x=5/9$, the $a$ lattice parameter increases, to reach a maximum value of 9.15~\AA\ for a full intercalated Na monolayer. This corresponds to an average percentage change of 0.7\% compared to the clean multilayers. 
This is in agreement with the overall increase in the in-plane lattice constant with Na intercalation predicted by Eames et al.~\cite{eames2014ion} and can be attributed to an increase in the lateral electrostatic repulsion between the Na atoms at increasing concentration. 

The change in the out-of-plane lattice constant upon Na intercalation is, unsurprisingly, significantly higher. The insertion of a single Na atom increases $c$ by an average value of 1.58~\AA, from 18.87~\AA\ to 20.45~\AA. The $c$ lattice constant then increases with increasing Na concentration, rising by 0.55~\AA\ at $x=5/9$, before decreasing slowly (by 0.2~\AA) with further increases in Na, to reach 20.80~\AA\ for a full intercalated monolayer.

\VC\ behaves broadly similarly upon Na intercalation, as shown in Fig.~\ref{fig:structure_changes}(b). The in-plane lattice constant initially decreases by a similar amount to \TiC, but the average maximum increase is slightly larger, at 0.16~\AA, corresponding to an increase of 1.8\%. 
In general, the spread of values across the individual mixed termination structures is slightly larger for \VC\ than for \TiC. 
The out-of-plane lattice constant also behaves similarly to \TiC\ after Na intercalation, especially for low concentrations of Na. There is a 1.60~\AA\ increase in $c$ after the intercalation of the first Na atom, from 13.7~\AA\ to 15.3~\AA. This rises to a maximum increase of 2.18~\AA\ for $x=4/9$ before decreasing slowly. This is an almost identical increase compared to the case of \TiC. The $c$ value starts to increase again for a full intercalated Na monolayer. 

Given that structural changes during charging and decharging are the primary cause of performance degradation in batteries, small volume changes after the first intercalation are desirable. \TiC\ and \VC\ have a maximum volume change of 11\% and 16\%, respectively, compared to the un-intercalated structure. However, some experiments have reported that Na does not always completely deintercalate from the MXene material, even at high voltages, i.e., it behaves as a pillar holding the interlayer distance open \cite{lukatskaya2013cation, kajiyama2016sodium} with further changes in the $c$ lattice constant restricted to $\pm$0.4\AA. We find the change in the $c$ lattice value reduces to 2.8\% and 4.2\%, for \TiC Na$_x$ and \VC Na$_x$, respectively, if we assume that after the first sodiation the interlayer spacing is held open by a single Na atom. These changes correspond to a change of $\pm$0.6~\AA\ for both \TiC Na$_x$ and \VC Na$_x$, in excellent agreement with experiment. 

\subsubsection{Comparison to Experiment}

A comparison between these lattice changes and the experimentally reported values should be performed carefully. Firstly, as stated previously, the composition of the terminating groups will affect the lattice constants, so that MXene created using different acids or different concentrations of acids will likely yield different changes in the lattice parameter. For example, Lukatskaya et al.~found that increases in the $c$ lattice constant depend sensitively on the type of salt used in the electrolyte. They showed that the $c$ value of \TiC\ increased by as much as 5.0~\AA\ in high-pH solutions such as Na$_2$CO$_3$, while neutral solutions such as the sulphates only increased $c$ by approximately 0.7~\AA~\cite{lukatskaya2013cation}. As such, the results presented here will only be valid for the stated composition of terminating groups. That being said, the spread of values indicated by the shaded regions in Fig.~\ref{fig:structure_changes} show how sensitive changes in the lattice parameter induced by Na intercalation are to changes in the terminating groups. 

Secondly, several experiments have reported the existence of a Na double layer intercalated between the layers~\cite{dall2015two, wang2015atomic}.
To determine if this is the case, we calculate the effects of intercalating a Na double layer in \TiC\ and \VC\ for a limited number of structures.
For both \TiC Na$_2$ and \VC Na$_2$, the $c$ lattice constant increases by 7.97~\AA\ compared to the unintercalated case. The vertical distance between the two Na layers is 3.1~\AA, which is similar to that calculated for a Na bilayer adsorbed on a Ti$_3$C$_2$O$_2$ monolayer (3.08\AA)~\cite{yu2016prediction}. The average vertical distance between the Na atoms and the terminating atoms is 1.5~\AA, while the average distance between Na and the carbon atoms is 3.93~\AA.
The increase in the $c$ lattice parameter found here for a Na double layer in \VC\ is considerably higher than the experimentally reported increase of 4.6~\AA~\cite{dall2015two}, suggesting that this value does not indicate the presence of a Na double layer, but rather other intercalants such as H$_2$O, or a large solvation shell around the intercalated Na atoms.

\subsubsection{Co-intercalated water}
As previously mentioned, many experimental structures also have considerable quantities of H$_2$O or other co-intercalants located between the MXene layers~\cite{hope2016nmr}. In many cases, drying the material in vacuum results in a reduction in the $c$ lattice parameter found initially, as the water is evicted. Kajiyama et al.~\cite{kajiyama2016sodium} show the $c$ lattice parameter of \TiC\ to increase from 19.4~\AA\ to 25~\AA\ upon first sodiation. However, after drying the material, the $c$ value decreases to 21.8~\AA, which is in much better agreement with the theoretical value of 20.80~\AA\ found here. Similarly, Osti et al.~\cite{osti2017influence} reports two peaks in the X-ray diffraction pattern of Na intercalated \TiC, corresponding to $c$ lattice constants of 23.98~\AA\ and 21.22~\AA. The smaller of these two values is in good agreement with our theoretical value, while the former could indicate the presence of co-intercalants. 

While a thorough investigation of the effects of water co-intercalation on the structural properties of these materials is beyond the scope of this work, an initial indication of its effect can be found by the introduction a single H$_2$O molecule into the \TiC Na$_{\,5/9}$ and \VC Na$_{\,5/9}$ structures. 
For the case of \TiC Na$_{\,5/9}$, we find that the $c$ lattice constant increases by 0.29~\AA, from 20.99~\AA\ to 21.28~\AA. For \VC Na$_{\,5/9}$, the lattice constant increases by 0.5~\AA, from 15.81~\AA\ to 16.32~\AA. 
The larger increase found for \VC\ compared to \TiC\ for the same concentration of co-intercalated water could be responsible for the unexpectedly large $c$ lattice constant of \VC~\cite{naguib2013new, harris2015direct, ghidiu2016ion}. 
In both cases, the introduction of H$_2$O causes Na atoms to move away from their high-symmetry adsorption position. This would agree with the experimental work of Shi et al.~who report that the intercalated Na atoms are disordered and do not sit at well defined locations~\cite{shi2014structure}.

\subsubsection{Electronic Properties}

\begin{figure}[htp]
\begin{centering}
\includegraphics[width=\linewidth]{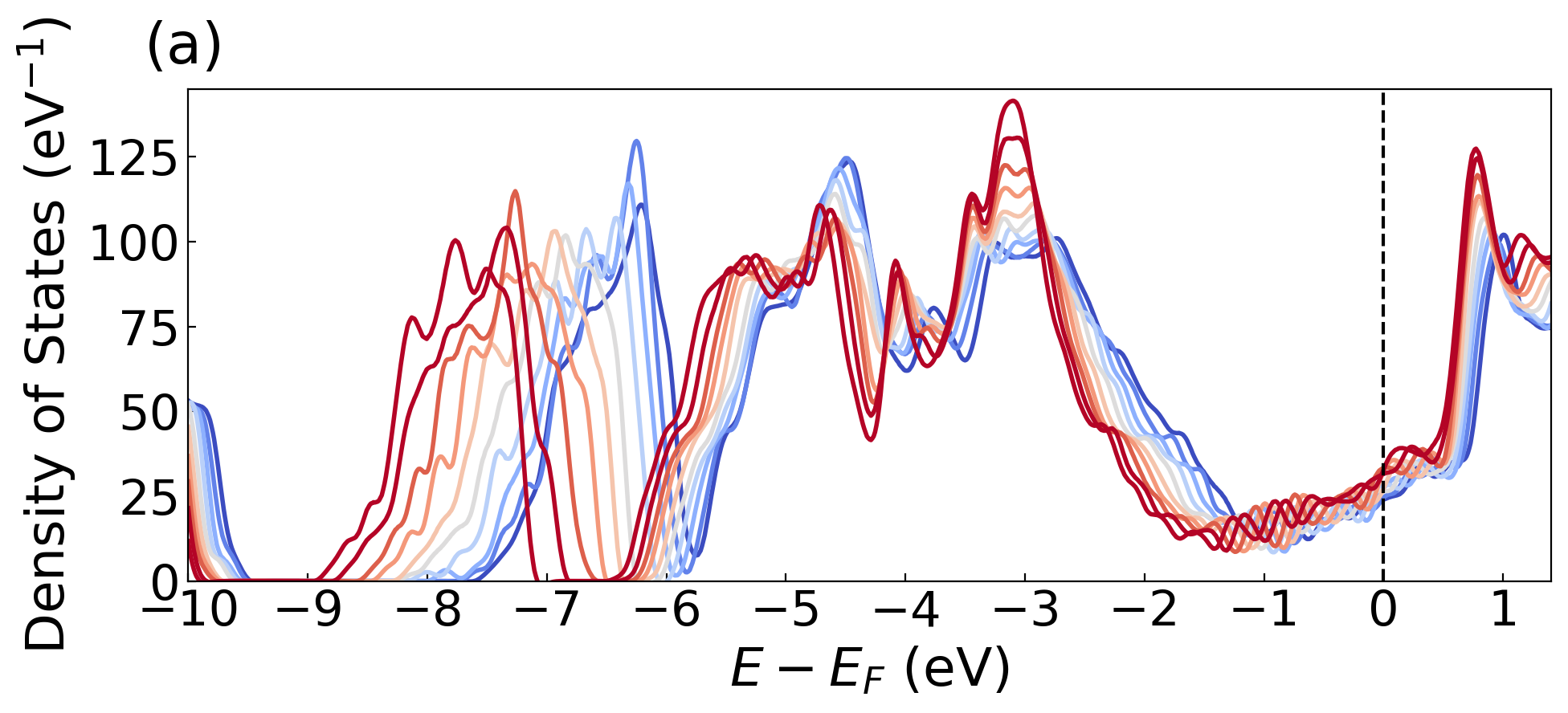}
\includegraphics[width=\linewidth]{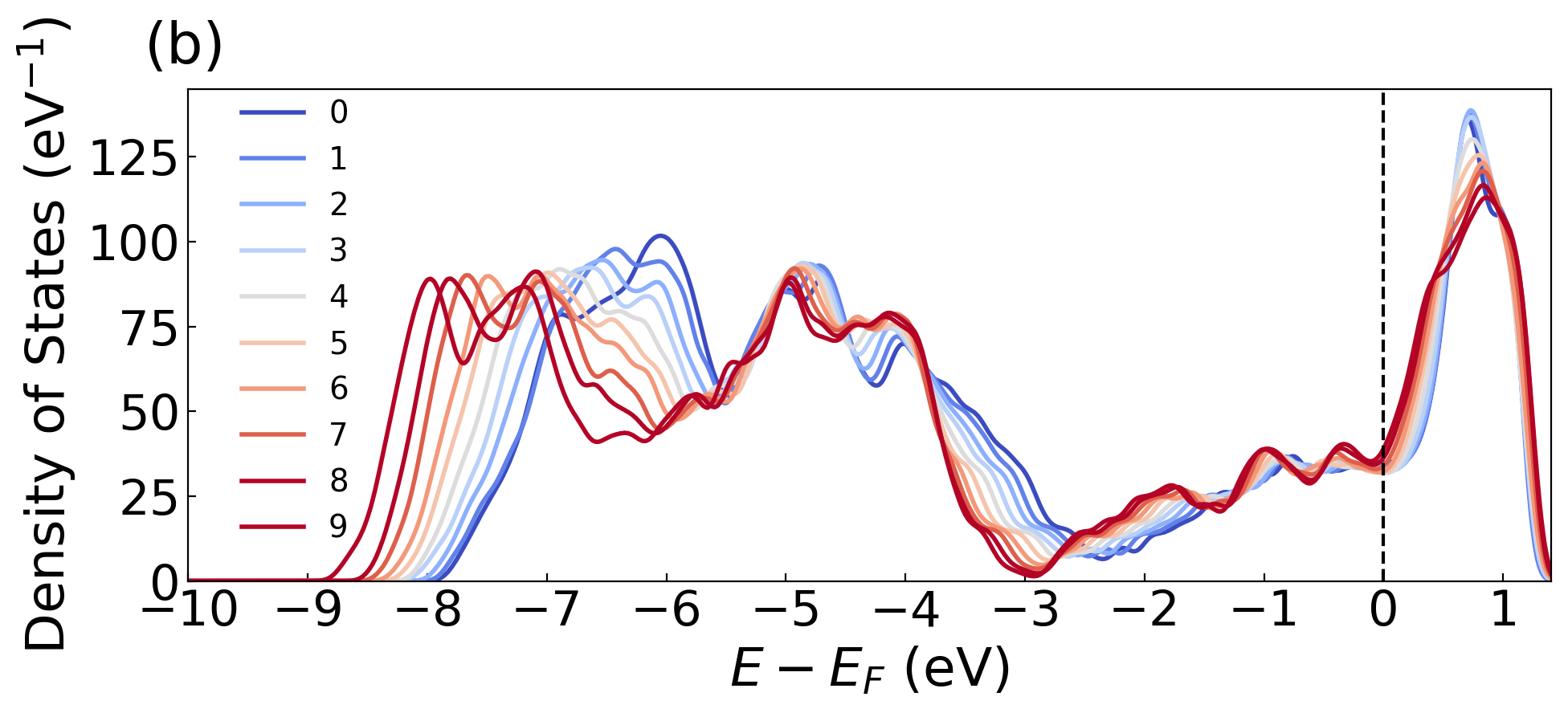}
\caption{\label{fig:na_dos} Total density of states of (a) Ti$_3$C$_2$T$_2$Na$_x$ and (b) V$_2$CT$_2$Na$_x$ as a function of Na intercalation, ranging from the clean stacked MXene ($x = 0$ dark blue) to \TiC Na ($(3 \times 3)x = 9$, dark red).}
\end{centering}
\end{figure}

The total density of states of stacked, mixed terminated \TiC\ and \VC\ are shown in Fig.~\ref{fig:na_dos} for increasing concentrations of intercalated Na. For each value of Na concentration, the DOS was averaged over the four considered mixed termination structures. By comparing to Fig.~\ref{fig:averaged_dos}(e) and (f), we can see that Na intercalation results in the states associated with the --F groups being moved to lower energies by approximately 1.3~eV in both materials. For the case of \TiC Na$_1$, this results in a gap of 0.7~eV opening between these F states and those associated with O and Ti. The states associated with O are also moved to lower energies by approximately 0.5~eV for a full intercalated Na monolayer compared to the unintercalated case in both materials. 

\begin{figure}[htp]
\begin{centering}
\includegraphics[width=\linewidth]{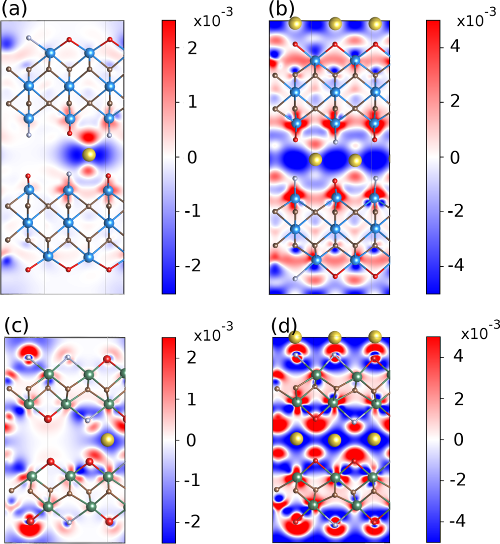}
\caption{\label{fig:charge_transfer} Slices of charge density difference after sodiation, shown in a plane perpendicular to the MXene layers and through the center of the $3 \times 3$ unit cell for \TiC Na$_x$, where (a) $x=1/9$, (b) $x=1$, and \VC Na$_x$, where (c) $x=1/9$, (d) $x=1$. Here, red and blue colors indicate accumulation and depletion of charge density, respectively. The scale is in units of a.u.$^{-3}$.}
\end{centering}
\end{figure}

To determine the effect of Na intercalation on the charge distribution, we calculated the change in charge density that occurs after intercalation. This charge density difference (CDD) is shown in Fig.~\ref{fig:charge_transfer} (a) and (b) for the intercalation of a single Na atom and a full Na monolayer, respectively, into the interlayer space of \TiC\ for a particular mixed termination. In both cases, there is a reduction of charge density around the Na atom. Bader charge analysis~\cite{henkelman2006fast} finds that, regardless of the concentration, the Na atom loses 0.8e, which is transfered to the MXene layer. For a single intercalated Na atom, this charge is predominantly located on the terminating groups nearest it. For the case shown here, the closest O atom gains approximately 0.09e, while the F atom gains 0.03e. Considerably less charge is transferred to the Ti atoms (0.03e is transferred to the Ti closest to the Na atom) and there is no charge transferred at all to the carbon atoms. For the case of a full intercalated Na monolayer, there is a large increase in the amount of charge transferred to the two extremal Ti layers. Each Ti gains 0.11e, so that their oxidation value rises from 2.35e to 2.46e. In contrast, the internal Ti layer and the C atoms gain very little excess charge (0.02e). The accumulation of charge on the terminating O atoms rises to 0.34e while the F atoms now gain 0.13e. This is in agreement with the experimental work of Kajiyama et al.~\cite{kajiyama2016sodium} who find that the redox reaction occurs mainly at the T$_\mathrm{x}$ groups, rather than at the Ti atoms. Lukatskaya et al.~\cite{lukatskaya2015probing} find that the oxidation state of the Ti atoms rises from 2.33e to 2.43e (i.e., by 0.1e per Ti atom) over a 0.7V window, in excellent agreement with the results presented here.

A very similar picture emerges for Na intercalation into \VC, as shown in Fig.~\ref{fig:charge_transfer} (c) and (d) for a single intercalating Na atom and a full Na monolayer, respectively. For the case of a single inserted Na atom, 0.8e is redistributed from the Na atom to the neighboring terminating atoms of the MXene layer. For a full intercalated Na monolayer, 0.8e from each of the Na atoms is again distributed to the O atoms (0.36e) and to the F atoms (0.16e). The V atoms gain 0.11e, so that their oxidation state rises from 3.52e to 3.63e. Bak et al.~\cite{bak2017ion} report that the electrochemical redox reaction mainly takes place at the vanadium sites during the sodiation process, with an average change in the vanadium oxidation state of 0.2e.

\subsubsection{Open Circuit Voltage and Specific Capacity}
The voltage delivered by a rechargeable battery is a key quantity to determine the effectiveness of a novel electrode material. 
Assuming the following half-cell reaction $vs$ Na/Na$^+$:
$$
\mathrm{Ti}_3\mathrm{C}_2\mathrm{T}_2 + x \mathrm{Na}^+ + xe^+ \longleftrightarrow \mathrm{Ti}_3\mathrm{C}_2\mathrm{T}_2\mathrm{Na}_x
$$
and neglecting volume and entropy effects, the open circuit voltage (OCV) can be approximated as~\cite{PhysRevB.56.1354}:
\begin{equation} \label{eq1}
\begin{split}
\mathrm{OCV} & \approx -\frac{E_f}{x} \\
 & = -\frac{1}{x} \left[E_{\mathrm{MXene}+x\mathrm{Na}} - E_{\mathrm{MXene}} - x E_{\mathrm{Na}}  \right] 
\end{split}
\end{equation}
where $E_f$ is the formation energy of sodiation, $x$ is the number of Na atoms intercalated between the layers of the stacked MXene structure and $E_{\mathrm{Na}}$ is the energy of metallic Na in its bulk body centered cubic structure. 
\begin{figure}[htp]
\begin{centering}
\includegraphics[width=\linewidth]{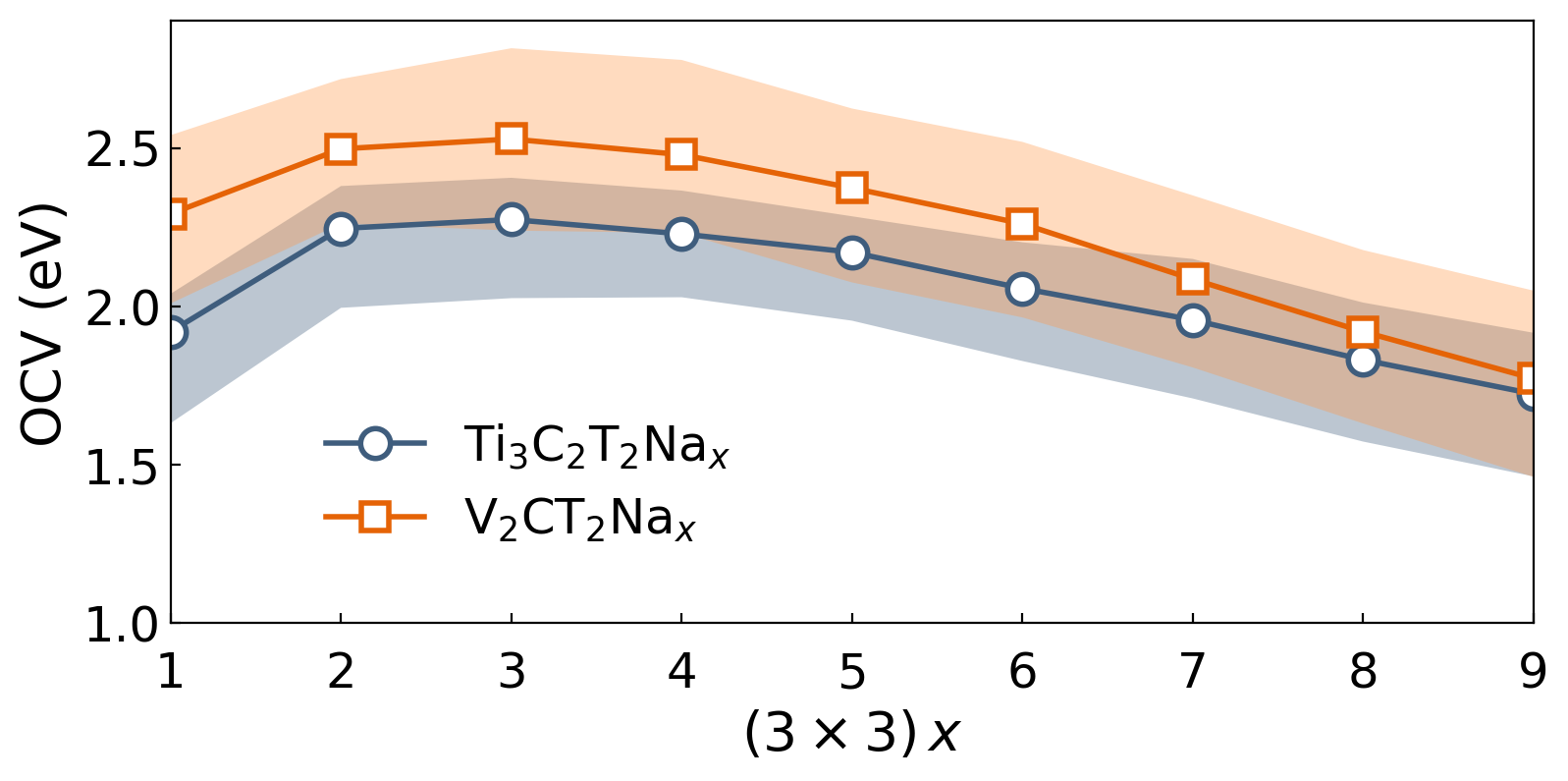}
\caption{\label{fig:ocv} OCV as a function of the number of intercalated Na atoms, $x$, in \TiC Na$_x$ (blue circles) and \VC Na$_x$ (orange squares). The solid line is an average over 4 different termination configurations which have values that range between the shaded areas.}
\end{centering}
\end{figure} 

The OCV is shown in Fig.~\ref{fig:ocv} for both \TiC Na$_x$ and \VC Na$_x$ with mixed terminations and for increasing concentrations of intercalated Na.
The OCV is positive over the entire range of Na concentrations considered and for both \TiC\ and \VC. While the OCV of \VC\ is higher by almost 0.4~eV at the lowest concentration of Na considered, the difference between the two materials decreases approximately linearly with increasing Na concentration to 0.05~eV. At this point, $x=1$, the OCV of both materials is approximately 1.7~eV.
In both cases, the OCV reaches a maximum at a  Na concentration of $x=1/3$. The shaded regions in Fig.~\ref{fig:ocv} indicates how sensitive the OCV is to the details of the individual termination configurations. For the case of \TiC, the OCV varies over an average range of approximately 0.39~eV, while for \VC\ the range is larger, at 0.54~eV.

Given that the OCV is related to the formation energy, the positive value at $x=1$ indicates that a full intercalated Na monolayer does not correspond to maximum capacity. We find the OCV associated with a Na double layer remains positive at 0.71~eV for \TiC Na$_2$ and 0.78~eV for \VC Na$_2$. 
While this would suggest that the concentration of intercalated Na atoms can be greater than even two monolayers, the experimentally reported capacity suggests that this is not achieved in practice.

The theoretical specific capacity is defined, in units of mAhg$^{-1}$, as: 
$$
C_t = \frac{xzF}{3.6 \left[ M_{\mathrm{electrode}} + xM_{\mathrm{Na}} \right] },
$$
where $x$ is the number of intercalated atoms, $z$ is the valance state of the intercalated atoms, in this case 1, $F$ is Faraday's constant, $M_{\mathrm{electrode}}$ is the molar mass of the electrode material and $M_{\mathrm{Na}}$ is the molar mass of a Na atom. 
\begin{figure}[ht!]
\begin{centering}
\includegraphics[width=0.95\linewidth]{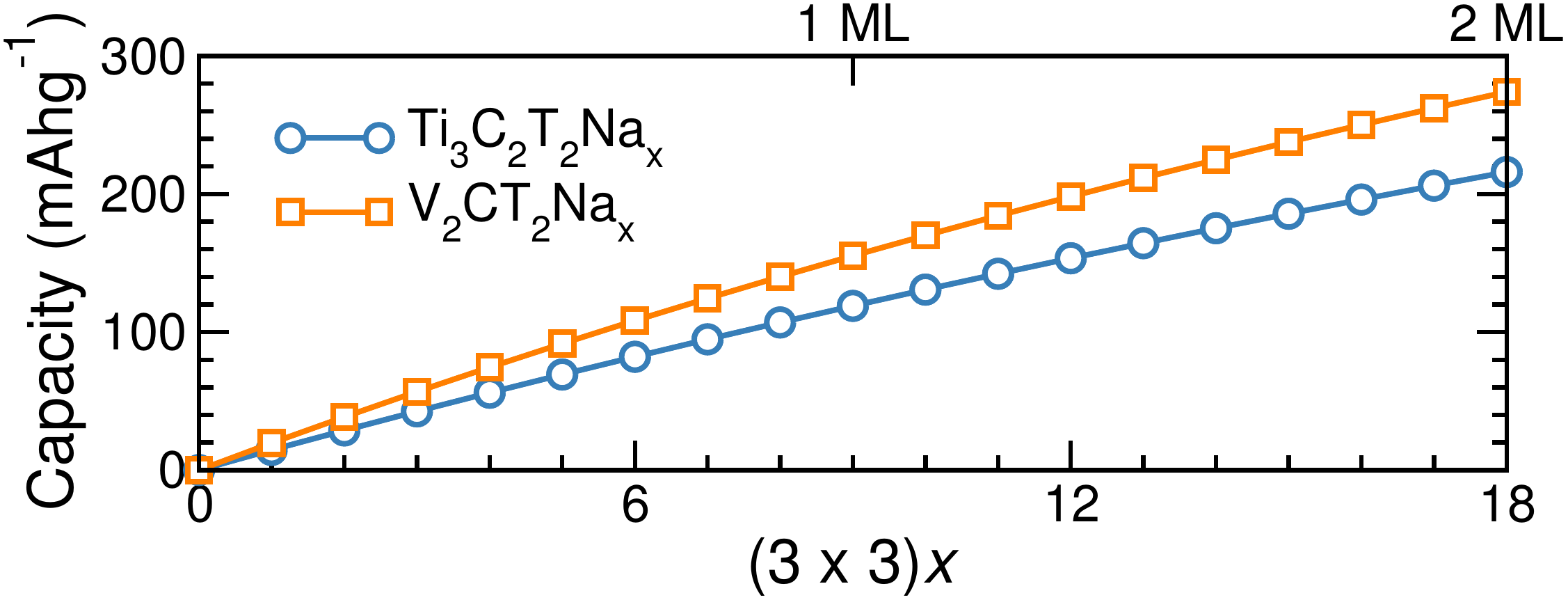}
\caption{\label{fig:capacity} Theoretical specific capacity of stacked \TiC Na$_x$ and \VC Na$_x$ using the molar mass of the structures with randomly populated terminations according to Table~\ref{tab:stoichiometry}.}
\end{centering}
\end{figure} 

The calculated specific capacities of \TiC Na$_x$ and \VC Na$_x$, are shown in Fig.~\ref{fig:capacity}. For the case of a full intercalated monolayer, the theoretical specific capacity of stacked \TiC Na$_1$ is 119~mAhg$^{-1}$, whereas it is 155~mAhg$^{-1}$ for the case of stacked \VC Na$_1$. Note here that the exact make-up of the terminating groups contributes relatively little to the total specific capacity, given the dominance of the molar weight of Ti and V to the total. The specific capacity of \TiC\ and \VC\ with a double layer of sodium atoms is 216~mAhg$^{-1}$ and 274~mAhg$^{-1}$, respectively. 
The higher specific capacity of \VC\ is as a result of its smaller molar mass compared to \TiC.

Xie et al.~report a first cycle discharge capacity of 370~mAhg$^{-1}$ for \TiC. The initial charge capacity is smaller at 164~mAhg$^{-1}$. By 10 cycles the capacity has reduced to 100~mAhg$^{-1}$ and finally to 80~mAhg$^{-1}$ at 120 cycles~\cite{xie2014prediction}. Kajiyama et al.~find a first sodiation capacity of 270~mAhg$^{-1}$, reducing to 100~mAhg$^{-1}$ at 100 cycles~\cite{kajiyama2016sodium}. While such high capacities for the first sodiation may suggest a very high intercalation concentration, this is not supported by structural data. Instead, such high capacities may be as a result of side reactions including the formation of a solid-electrolyte interphase (SEI) or electrolyte decomposition. A capacity of approximately 100~mAhg$^{-1}$ would correspond to a Na concentration of $x=0.9$. 
Bak et al.~report an experimental capacity of 200~mAhg$^{-1}$ during the first sodiation discharge in \VC, and a capacity of 125~mAhg$^{-1}$ for the first charge. This is reduced to 90~mAhg$^{-1}$ after the first few cycles~\cite{bak2017ion}.
Similarly, Dall'Agnese et al.~\cite{dall2015two} report an specific capacity of approximately 70~mAh/g at low scanning rates for \VC. This would correspond to a Na concentration of $x=0.44 - 0.55$ in \VC Na$_x$.
Many factors can result in the experimental capacity being lower that the predicted value, including solvation effects, the presence of solvated ions and the presence of any reside of the precursor ternary carbide material.

\section{Conclusion}

We have examined how mixed termination groups affect the structural, electronic and electrochemical properties of \TiC\ and \VC. 
Several structures with mixed terminations were considered such that their average reflected a typical experimental composition. We show that while these individual structures show the properties to be quite sensitive to the particular terminating groups, the average value is generally in very good agreement with experiment. 

Furthermore, in cases where taking mixed terminations into account explicitly is unfeasible due to the computational expense, a significantly improved approximation can be achieved by taking a suitable weighted average of uniformly-terminated layers.
This procedure was shown to yield accurate results for the lattice parameters, the electronic density of states and the work function of these materials.

Interestingly, we highlight that some of the highly desirable properties postulated for uniformly terminated surface are not highly sensitive to the exact surface termination.
For example, the ultra-low work functions proposed for uniformly --OH terminated MXenes can be achieved with only 60\% --OH groups on the surface.
Such a result demonstrates that, despite the experimental difficulties in controlling the surface terminations of these MXene materials, they remain highly promising for technological applications.

Finally, the sodium storage capacity and structural changes during sodiation were also investigated for the mixed terminated \TiC\ and \VC\ structures. The redox reaction was found to be confined to the terminating groups at low concentrations of intercalated Na, with the oxidation state of the metal atoms unaffected until a higher Na concentration is achieved, in excellent agreement with experiment. The open circuit voltage is shown to be highly sensitive to the particular composition of the terminating groups, varying by approximately 0.4~eV at each Na concentration. 

\section*{Conflicts of interest}
There are no conflicts to declare.

\section*{Acknowledgements}
This work was supported by a Science Foundation Ireland Starting Investigator Research Grant (15/SIRG/3314). 
Computational resources were provided by the supercomputer facilities at the Trinity Center for High Performance Computing (TCHPC) and at the Irish Center for High-End Computing (project tcphy091b). 

\bibliography{mxene} 
\bibliographystyle{rsc}

\end{document}